\documentclass[10pt]{iopart}
\usepackage{color}
\usepackage{textcomp}
\usepackage{bm}
\usepackage{cuted}
\usepackage{braket}
\expandafter\let\csname equation*\endcsname\relax
\expandafter\let\csname endequation*\endcsname\relax
\usepackage{amsmath}
\usepackage{amssymb}
\usepackage{upgreek}
\usepackage{multirow}
\usepackage{svg}
\usepackage{graphicx}
\usepackage{hyperref}
\hypersetup{colorlinks=true, citecolor=blue, urlcolor=blue, linkcolor=blue}
\usepackage[sort&compress,square,numbers]{natbib}

\begin{document}

\title{Three Dimensional Orientation of Complex Molecules Excited by Two-Color Femtosecond Pulses}
\author{Long Xu$^1$, Ilia Tutunnikov$^1$, Yehiam Prior, Ilya Sh. Averbukh}
\address{AMOS and Department of Chemical and Biological Physics, The Weizmann Institute of Science, Rehovot 7610001, Israel}
\address{$^1$These authors contributed equally to this work.}
\ead{long.xu@weizmann.ac.il, ilia.tutunnikov@weizmann.ac.il, yehiam.prior@weizmann.ac.il, and ilya.averbukh@weizmann.ac.il}

\begin{abstract}

We study the excitation of asymmetric-top (including chiral) molecules by two-color femtosecond laser pulses.
In the cases of non-chiral asymmetric-top molecules excited by an orthogonally polarized two-color pulse, we demonstrate, classically and quantum mechanically, three-dimensional orientation. For chiral molecules, we show that the orientation induced by a cross-polarized two-color pulse is enantioselective along the laser propagation direction, namely, the two enantiomers are oriented in opposite directions. On the short time scale, the classical and quantum simulations give results that are in excellent agreement, whereas on the longer time scale, the enantioselective orientation exhibits quantum beats.
These observations are qualitatively explained by analyzing the interaction potential between the two-color pulse and molecular (hyper-)polarizability. The prospects for utilizing the long-lasting orientation for measuring and using the enantioselective orientation for separating the individual enantiomers are discussed.

\end{abstract}


\maketitle
\ioptwocol

\section{Introduction} \label{sec:Introduction}

Molecular alignment and orientation have attracted widespread interest due to their importance in chemical reaction control, ultra-fast imaging of molecular structure and dynamics, and photon-induced molecular processes. Notable examples, to name a few, are the utilization of ultra-short laser pulses to induce field-free three-dimensional (3D) alignment of gas-phase molecules \cite{Larsen2000Three,Underwood2005,Lee2006Field,Artamonov2008Theory,Ren2014,Saribal2021}, and of small molecules trapped in helium droplets, e.g., see \cite{Chatterley2019}; or the use of terahertz (THz) pulses to prepare oriented samples of polar symmetric- \cite{Babilotte2016Observation,Xu2020} and asymmetric-top \cite{Babilotte2017,Damari2017Coherent} molecules under field-free conditions. For extensive reviews of the molecular alignment and orientation, see \cite{StapelfeldtSeidman2003,Ohshima2010,Fleischer2012,Lemeshko2013,Koch2019Quantum,Lin2020Review}. An additional efficient  tool for inducing molecular orientation is non-resonant phase-locked two-color laser pulses consisting of the fundamental wave (FW) and its second harmonic (SH). Two-color pulses have been shown, both theoretically and experimentally, to be effective for orienting linear molecules \cite{Vrakking1997,Dion1999,Kanai2001,De2009,Oda2010,JW2010,Zhang2011-multicolor,Frumker2012,Spanner2012,Znakovskaya2014,Mun2018,Mun2019,Mun2020,MelladoAlcedo2020,Shuo2020}. Recently, we theoretically considered the orientation of symmetric-top molecules excited by femtosecond two-color pulses \cite{xu2021longlasting}, and previously experimentally demonstrated field-free 3D orientation induced by orthogonally polarized femtosecond two-color pulses \cite{Lin2018All}.

A unique class of asymmetric-top molecules is formed by the chiral molecules. A chiral molecule has two forms, called left- and right-handed enantiomers. Control schemes enabling manipulating specific enantiomers are called enantioselective. Following a series of theoretical works \cite{Yachmenev2016,Gershnabel2018,Tutunnikov2018,Tutunnikov2019Laser}, the enantioselective orientation of chiral molecules was demonstrated experimentally \cite{Milner2019Controlled,Tutunnikov2020Observation} using the optical centrifuge for molecules \cite{Karczmarek1999Optical,Villeneuve2000Forced,Yuan2011Dynamics,Korobenko2014Direct,Korobenko2018Control}. Most recently, THz pulses with twisted polarization were theoretically shown to induce enantioselective orientation as well \cite{TutunnikovXu2020}.

This paper considers the orientation of asymmetric-top (including chiral) molecules excited by two-color femtosecond laser pulses. We begin by analyzing molecules belonging to the $C_{2v}$ point symmetry group, including formaldehyde and sulfur dioxide as typical examples, excited by orthogonally polarized two-color pulses. In these cases, the two-color pulse excitation gives rise to both 3D alignment and orientation of the permanent molecular dipole moment. Then, we investigate the orientation of chiral molecules (lacking any symmetry) excited by pulses with general cross-polarization, where the FW and SH polarizations are neither parallel nor orthogonal. In this case, the induced orientation is shown to be \emph{enantioselective}. The paper is organized as follows: in the next section, we summarize our numerical methods. In  Sec. \ref{sec:3Dalignment}, we present the three-dimensional molecular alignment/orientation and the long-lasting orientation induced by the orthogonally polarized two-color pulse. In Sec. \ref{sec:enantioselective-orientation-effect}, we discuss  the enantioselective orientation of chiral molecules, and finally, Sec. \ref{sec:Conclusion} concludes the paper.

\begin{table*}[!t]
\caption{\label{tab:Molecular-properties}Molecular properties (in atomic units) of  $\mathrm{CH_2O}$ and $\mathrm{SO_2}$: moments of inertia,
 non-zero elements of dipole moment, polarizability tensor, and hyperpolarizability tensor in the frame of molecular principal axes of inertia. }
\begin{indented}
\item[]
\begin{tabular}{@{}ccccc}\br
 Molecules                  & Moments of inertia            & Dipole
components           & Polarizability components                    & Hyperpolarizability components                    \\
\mr
\multirow{3}{2.2cm}{{Formaldehyde $\mathrm{CH_2O}$}} & {$I_{a}=11560$} & {$\mu_{a}=-0.948$}& {$\alpha_{aa}=22.64$}& {$\beta_{aaa}=47.0$}\tabularnewline
 & {$I_{b}=84122$} & & {$\alpha_{bb}=18.16$}& {$\beta_{abb}=62.9$}\\
 & {$I_{c}=95682$} & & {$\alpha_{cc}=13.06$}& {$\beta_{acc}=10.3$}\\
\mr
\multirow{3}{2.2cm}{{Sulfur dioxide $\mathrm{SO_2}$}} & {$I_{a}=55509$} & & {$\alpha_{aa}=31.26$}& {$\beta_{aab}=22.0$}\\
 & {$I_{b}=317477$} & {$\mu_{b}=-0.7877$}& {$\alpha_{bb}=20.80$}& {$\beta_{bbb}=26.5$}\\
 & {$I_{c}=371885$} & & {$\alpha_{cc}=18.64$}& {$\beta_{bcc}=6.4$}\\
 \br
\end{tabular}
\end{indented}
\end{table*}

\section{Numerical methods} \label{sec:Numerical methods}

We consider the asymmetric-top molecules as rigid rotors. The electric field of two-color laser pulses, consisting of the FW and its SH, is modeled by
\begin{equation}
\mathbf{E}(t) = \varepsilon_1(t)\cos(\omega t)\mathbf{e}_X +  \varepsilon_{2}(t)\cos(2\omega t)\mathbf{e}_{\mathrm{SH}}
,\label{eq:electric-field-lab-frame}
\end{equation}
where $\varepsilon_{n}(t)= \varepsilon_{n,0}\exp[-2\ln 2\, (t/\sigma_n)^2],\, n=1,2$, is the field's envelope, $\varepsilon_{n,0}$ is the peak amplitude, and $\sigma_n$ is the full width at half maximum (FWHM).
Here, $\omega$ is the carrier frequency of the FW field, $\mathbf{e}_{\mathrm{SH}} = \cos(\phi_{\mathrm{SH}})\mathbf{e}_X + \sin(\phi_{\mathrm{SH}})\mathbf{e}_Y$,
where $\phi_{\mathrm{SH}}$ is the angle between the polarizations of the FW and SH fields,
and $\mathbf{e}_{X}$ and $\mathbf{e}_{Y}$ are unit vectors along the laboratory $X$ and $Y$ axes, respectively.

Simulations of the laser-driven molecular rotations were carried out both classically and fully quantum mechanically. An extended description of our numerical approaches can be found in \cite{xu2021longlasting}.

\subsection{Classical simulation} \label{sec:Classical simulation}

The rotation of a classical rigid body is described by Euler's equations \cite{Goldstein2002Classical}
\begin{equation}
\mathbf{I}\bm{\dot{\Omega}}=(\mathbf{I}\bm{\Omega})\times\bm{\Omega}+\mathbf{T},\label{eq:Eulers-equations}
\end{equation}
where $\mathbf{I}=\mathrm{diag}(I_{a},I_{b},I_{c})$ is the moment of inertia tensor (here, $I_a<I_b<I_c$), $\bm{\Omega}=(\Omega_{a},\Omega_{b},\Omega_{c})$ is the angular velocity, and $\mathbf{T}=(T_{a},T_{b},T_{c})$ is the external torque, $x\times y$ denotes vector cross-product. All quantities in Eq. \eqref{eq:Eulers-equations} are expressed in the molecular reference frame where the basis comprises the three principal axes of inertia, $a,\,b,\,c$. When external electric fields couple to a rigid body via its polarizability and hyperpolarizability, the torque is given by
\begin{align}
\mathbf{T} &= \bm{\alpha}\mathbf{E}\times \mathbf{E}+\frac{1}{2}\left(\mathbf{E}\bm{\beta} \mathbf{E}\right)\times \mathbf{E},\label{eq:torque}
\end{align}
where $\bm{\alpha}$ and $\bm{\beta}$ are the polarizability and hyperpolarizability tensors, respectively, and $\mathbf{E}$ is the electric field vector.
The time-dependent relation between the laboratory and molecular reference frames is parametrized by a quaternion, $q$ \cite{Coutsias2004The,Kuipers1999Quaternions}.  Quaternions obey the equation of motion $\dot{q}=q\Omega/2$, where $\Omega$ is a quaternion composed of the molecule's angular velocity components \cite{Coutsias2004The,Kuipers1999Quaternions}.
To simulate the behavior of a thermal ensemble consisting of $N\gg1$ non-interacting molecules, we use the Monte Carlo method. For each molecule, we numerically solve a system of coupled equations consisting of Euler's equations [Eq. \eqref{eq:Eulers-equations}], including the torque in Eq. \eqref{eq:torque}, and the equation of motion of the quaternion.
The initial state, in which the molecules are isotropically distributed in space, is modeled using random uniform quaternions generated according to the recipe in \cite{Lavalle2006Planning}.
Initial molecular angular velocities are given by the Boltzmann distribution,
\begin{equation}
f(\bm{\Omega})\propto
\prod_{i}\exp\left(-\frac{I_{i}\Omega_{i}^{2}}{2k_{B}T}\right),\label{eq:Classical-Boltzmann-distribution}
\end{equation}
where $i=a,b,c$, $k_{B}$ is the Boltzmann constant, and $T$ is the temperature.

\subsection{Quantum simulation} \label{sec:quantum simulation}

The Hamiltonian describing the rotational dynamics of a molecule driven by two-color fields is $H(t)=H_{r} + H_\mathrm{int}(t)$, where $H_{r}$ is the field-free Hamiltonian \cite{zare1988Angular} and the molecule-field interaction potential, including the polarizability and hyperpolarizability contributions, is given by \cite{Buckingham2007Permanent}
\begin{equation}\label{eq:potential}
H_\mathrm{int}(t) = -\frac{1}{2} \sum_{i j} \alpha_{i j} E_{i} E_{j} -\frac{1}{6} \sum_{i j k} \beta_{i j k} E_{i} E_{j} E_{k}.
\end{equation}
Here $E_{i}$, $\alpha_{i j}$, and $\beta_{ijk}$ represent the components of the field vector, polarizability tensor, and hyperpolarizability tensor, respectively.

For the calculations, we use the basis set of field-free symmetric-top wave functions $|JKM\rangle$ \cite{zare1988Angular}.
Here $J$ is the total angular momentum, $K$ and $M$ are the projections onto the molecule-fixed $a$ axis and the laboratory-fixed $Z$ axis, respectively.
The non-zero matrix elements of the field-free Hamiltonian of asymmetric-top molecules are given by \cite{zare1988Angular}
\begin{flalign}\label{eq:HR}
&\langle JKM|H_{r}|JKM\rangle=\frac{B+C}{2}  \left[J(J+1)-K^{2}\right]+AK^{2}, \nonumber\\
&\langle JKM|H_{r}|JK\pm2M\rangle  =\frac{B-C}{4}f(J,K\pm1),
\end{flalign}
where
$ f(J,K)=\sqrt{(J^{2}-K^{2})[(J+1)^{2}-K^{2}]} $
and the rotational constants are $A=\hbar^{2}/2I_{a}$, $B=\hbar^{2}/2I_{b}$, and $C=\hbar^{2}/2I_{c}$ with $A>B>C$.

The eigenfunctions of the asymmetric top are obtained numerically by diagonalizing the field-free Hamiltonian $H_{r}$. The eigenfunctions can be written as $|J\tau M\rangle=\sum_{K}c_{K}^{(J,\tau,M)}|JKM\rangle$ [see Eq. \eqref{eq:HR}], where $-J\leq\tau\leq J$ enumerates the asymmetric-top eigenstates for a given $J,M$ quantum numbers.
We set the initial state as $|\Psi(t=0)\rangle=|J\tau M\rangle$ and solve the time-dependent Schr\"{o}dinger equation $i\hbar \partial_t |\Psi(t)\rangle = H(t)|\Psi(t)\rangle$ using numerical exponentiation of the Hamiltonian matrix (see Expokit \cite{sidje1998Expokit}).
The expectation value of the dipole signal along a direction defined by $\mathbf{e}_{i}$, a unit vector in the laboratory-fixed frame, is given by
\begin{align}\label{eq:Polarization}
\begin{split}
& \braket{\mu_{i}^{(J,\tau,M)}}\!(t)  =\langle\Psi(t)|\bm{\mu}\cdot\mathbf{e}_{i}|\Psi(t)\rangle,
\end{split}
\end{align}
where $\bm{\mu}$ is the molecular dipole moment vector.

\begin{figure}[!t]
\centering{}\includegraphics[width=\linewidth]{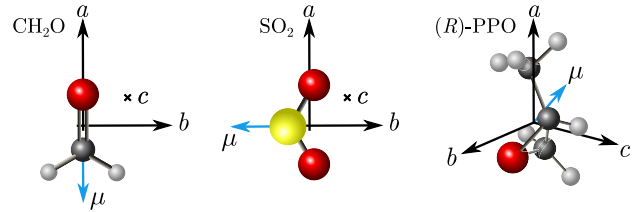}
\caption{Formaldehyde ($\mathrm{CH_2O}$), sulfur dioxide ($\mathrm{SO_2}$), and (\emph{R})-PPO molecules. Axes $a$, $b$, and $c$ are the principal axes of inertia tensor. Atoms are color-coded: black - carbon, gray - hydrogen, red - oxygen, sulfur - yellow. The light blue arrows represent the dipole moments. Note that the dipole moment of PPO has non-zero projections along all three molecular axes (see Table \ref{tab:Molecular-properties-PPO}).
 \label{fig:molecules}}
\end{figure}

Thermal effects are included by averaging the expectation value over the different initial states $|J\tau M\rangle$ with the relative weight given by the Boltzmann distribution,
\begin{equation}
\braket{\mu_{i}}\!(t) =\frac{1}{\mathcal{Z}}\sum\limits _{J,\tau,M}\exp\left(-\frac{\varepsilon_{J,\tau,M}}{k_{B}T}\right)\braket{\mu_{i}^{(J,\tau,M)}}\!(t) ,\label{eq:Quantum-Boltzmann-distribution}
\end{equation}
where $\mathcal{Z}=\sum_{J,\tau,M}\exp\left(-\varepsilon_{J,\tau,M}/k_{B}T\right)$,
and $\varepsilon_{J,\tau,M}$ is the energy of $|J\tau M\rangle$.

\section{Three-dimensional alignment and orientation}  \label{sec:3Dalignment}

In this section, we consider two relatively simple (non-chiral) asymmetric-top molecules, formaldehyde ($\mathrm{CH_2O}$)
and sulfur dioxide ($\mathrm{SO_2}$), excited by an orthogonally polarized two-color pulse. The polarizations of the FW and SH fields are set along the $X$ and $Y$ axes, respectively.
Molecular properties of $\mathrm{CH_2O}$ and $\mathrm{SO_2}$ are summarized in
Table \ref{tab:Molecular-properties}.
Moments of inertia are taken from NIST (DFT, method of CAM-B3LYP/aug-cc-pVTZ) \cite{johnson1999nist}, while the other parameters of $\mathrm{CH_{2}O}$ were taken from \cite{Benkova2007} (DFT, method of B3LYP/aug-cc-pVTZ).
The parameters of $\mathrm{SO_2}$ are taken from the literature \cite{Maroulis1992}.
Figure \ref{fig:molecules} presents graphical images of the molecules with their principal axes of inertia.

Since $\mathrm{CH_2O}$ and $\mathrm{SO_2}$ are planar molecules belonging to the $C_{2v}$ point symmetry group [see Fig. \ref{fig:molecules}], according to spin-statistics theorem, additional spin statistical factor should be included in the sum in Eq. \eqref{eq:Quantum-Boltzmann-distribution}.
In the case of $\mathrm{CH_2O}$ molecule, initial rotational $\ket{J\tau M}$ states formed by $\ket{JKM}$ states with an odd quantum number $K$ (antisymmetric with respect to $\pi$-rotation about the $a$ axis) have triple the statistical weight compared with $\ket{J\tau M}$ states formed by symmetric-top states with even $K$ (symmetric with respect to $\pi$-rotation about the $a$ axis). For details, see the introduction of \cite{BECHTEL2005Nuclear}.
For $\mathrm{SO_2}$ molecule, only states symmetric with respect to $\pi$-rotation about the $b$ axis (dipole moment) are taken into account \cite{Damari2016}.

\subsection{Formaldehyde}  \label{sec:formaldehyde}

\begin{figure}[!t]
\centering{}\includegraphics[width=\linewidth]{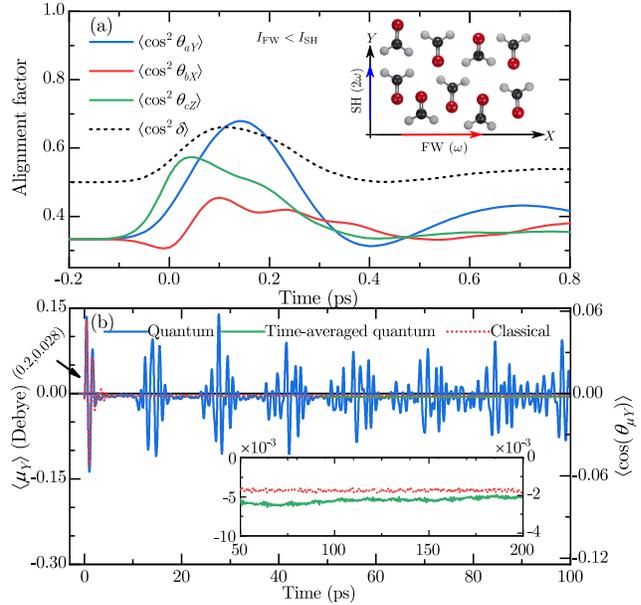}
\caption{(a) Classically calculated degrees of alignment for $\mathrm{CH_2O}$ molecule.
The overall degree of alignment is defined as $\braket{\cos^2\delta} = \left(1+\braket{\cos^2\theta_{aY}}+\braket{\cos^2\theta_{bX}}+\braket{\cos^2\theta_{cZ}}\right)/4$ (dashed black).
The inset illustrates the 3D alignment and orientation along the SH direction (notice, more oxygen atoms point down).
(b) $Y$ projection of the dipole signal.
Solid blue and dotted red lines are the results of quantum and classical simulations, respectively.
Solid green line is the moving time average defined by $\overline{\langle\mu_{Y}\rangle(t)}=(\Delta t)^{-1}\int_{t-\Delta t/2}^{t+\Delta t/2}\mathrm{d}t'\langle\mu_{Y}\rangle(t')$,
where $\Delta t=100\,\mathrm{ps}$. The inset shows a magnified portion of the signals.
Here, initial temperature is $T=5\,\mathrm{K}$, the angle between the polarizations of FW and SH is $\phi_\mathrm{SH} = \pi/2$. $N=10^8$ molecules are used in the classical simulation.
 \label{fig:CH2O}}
\end{figure}

Figure \ref{fig:CH2O}(a) shows the classically calculated degrees of alignment of $\mathrm{CH_2O}$ molecules during and shortly after the excitation by a femtosecond two-color pulse. The degrees of alignment are quantified by averages of squares of directional cosines, $\cos^2(\theta_{ij})$, where $\theta_{ij}$ represents the angle between the molecular $i$ axis and the laboratory $j$ axis.
Here, the initial temperature is $T= 5\, \mathrm{K}$, the peak intensities of the FW and SH fields are $I_\mathrm{FW}=2\times 10^{13} \,\mathrm{W/cm^2}$ and $I_\mathrm{SH}=8\times 10^{13} \,\mathrm{W/cm^2}$, the duration is $\sigma_1 =\sigma_2 = 120 \,\mathrm{fs}$, and $\phi_\mathrm{SH}=\pi/2$ [see Eq. \eqref{eq:electric-field-lab-frame}].
As can be seen, shortly after the pulse (at $t\simeq 0.2 \, \mathrm{ps}$), all three molecular axes are simultaneously aligned along the three laboratory axes.
The molecular $a$ axis is aligned along the laboratory $Y$ axis,
while the $b$ and $c$ axes are aligned along the $X$ and $Z$ axes, respectively.
Similar 3D alignment was achieved using a single one-color elliptically polarized laser pulse \cite{Larsen2000Three,Artamonov2008Theory,Chatterley2019,Saribal2021}, or a pair of delayed cross-polarized one-color pulses \cite{Underwood2005,Lee2006Field}. In the case of two-color pulses used here, the strong SH field aligns the most polarizable $a$ axis along the laboratory $Y$ axis [see the oxygen atom (red) in the inset of Fig. \ref{fig:CH2O}(a)], while the weak FW field aligns the second most polarizable $b$ axis (the intermediate axis of inertia) along the $X$ axis [see hydrogen atoms (gray) in the inset of Fig. \ref{fig:CH2O}(a)]. Finally, due to the geometric constraints, the $c$ axis is aligned along the $Z$ axis, resulting in the 3D alignment.

Moreover, Fig. \ref{fig:CH2O}(b) shows the calculated (both classically and quantum mechanically) expectation value of the projection of the dipole moment on the laboratory $Y$ axis, $\braket{\mu_{Y}}(t)$. The degrees of orientation of all the molecular axes along the $X$ and $Z$ axes are identically zero. Both the quantum and classical results show the transient dipole orientation along the $-Y$ direction following the kick by the two-color femtosecond pulse.

\begin{figure}[!t]
\centering{}\includegraphics[width=\linewidth]{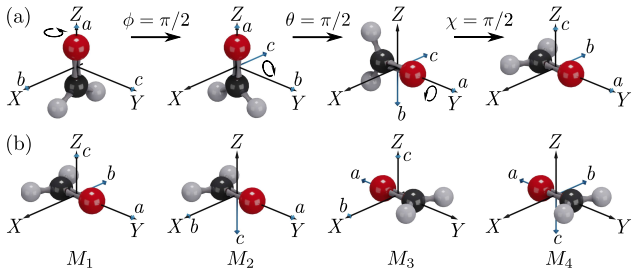}
\caption{(a) A sequence of three Euler rotations (according to convention of \cite{zare1988Angular}) bringing the  molecule-fixed frame ($bca$) into the orientation corresponding to one of the four global minima, $M_1=(\pi/2,\pi/2,\pi/2)$.
(b) Orientations of the molecule-fixed frame defined by four global minima, $M_i$. The dipole moment is $\boldsymbol{\mu}\propto-\bm{a}$.
\label{fig:3D_orientation}}
\end{figure}

Both the 3D alignment and orientation of the dipole moment can be understood by analyzing the field-polarizability interaction potential [see Eq. \eref{eq:potential}], which can be expressed in terms
of the three Euler angles $\phi,\,\theta,\,\chi$. Here, we adopt the convention used in \cite{zare1988Angular}, according to which $\phi$ and $\theta$ are the azimuthal and polar angles of the molecule-fixed $z$ axis ($a$ axis used here) with respect to the laboratory-fixed $X$ and $Z$ axes, respectively, and $\chi$ represents the additional rotation angle around the $z$ axis [see Fig. \ref{fig:3D_orientation}(a)].
For the orthogonally polarized two-color pulse used here, after averaging over the optical cycle, the field-polarizability interaction potential breaks into two independent parts [see Eq. \eqref{eq:potential_polarizability}]: (i) the interaction with the FW field, and (ii) the interaction with the SH field.

In the case of $\mathrm{CH_2O}$ molecule, the field-polarizability interaction potential [see Eq. \eqref{eq:potential_polarizability}] has four global minima at
$M_i=(\phi_i,\theta_i,\chi_i)$, where
$M_1=(\pi/2,\pi/2,\pi/2)$, $M_2= (\pi/2,\pi/2, 3\pi/2)$, $M_3=(3\pi/2,\pi/2,\pi/2)$, $M_4= (3\pi/2,\pi/2,3\pi/2)$.
Taking $M_1$ as an example, Fig. \ref{fig:3D_orientation}(a) shows the sequence of rotations bringing the molecule-fixed frame into the orientation defined by $M_1$.
All four orientations of the molecule-fixed frame are shown in Fig. \ref{fig:3D_orientation}(b), which together produce 3D alignment as illustrated in the inset of Fig. \ref{fig:CH2O}(a).

In terms of the transformation relating the laboratory-fixed and molecule-fixed frames of reference, $U(\phi,\theta,\chi)(b,c,a)^\mathrm{T}=(X,Y,Z)^\mathrm{T}$ [see Eq. \eqref{eq:transformation}], for the molecular dipole moment, the four minima satisfy
\begin{subequations}
\begin{align*}
&U(M_1)\bm{\mu}^\mathrm{T}=U(M_2)\bm{\mu}^\mathrm{T}=(0,-0.948,0)^\mathrm{T},\\ &U(M_3)\bm{\mu}^\mathrm{T}=U(M_4)\bm{\mu}^\mathrm{T}=(0,+0.948,0)^\mathrm{T},
\end{align*}
\end{subequations}
which means that the molecular dipole moment ($\boldsymbol{\mu}\propto-\bm{a}$) is \emph{aligned} along the laboratory $Y$ axis, see Fig. \ref{fig:3D_orientation}(b).
Similarly, it can be shown that the $b$ and $c$ axes are aligned along the $X$ and $Z$ axes, respectively.

Unlike the symmetric field-polarizability interaction, which has two independent contributions from the FW and SH, the field-hyperpolarizability interaction potential couples the FW and SH fields. This results in an asymmetric interaction potential causing \emph{orientation} of the dipole moment.
The field-hyperpolarizability interaction potential [see Eq. \eqref{eq:potential_CH2O}] has two global minima at $M_1$ and $M_2$.
Since
$$U(M_1)\bm{\mu}^\mathrm{T}=U(M_2)\bm{\mu}^\mathrm{T}=(0,-0.948,0)^\mathrm{T},$$
the symmetry along the laboratory $Y$ axis is broken, resulting in an imbalance of the dipole moments pointing along the $Y$ direction versus those pointing along the $-Y$ direction [see Fig. \ref{fig:3D_orientation}(b)]. Notice that the orientation is along the polarization direction of the SH field. For molecules belonging to the $C_{2v}$ point symmetry group, the combination of 3D alignment and orientation of one of the molecular axes results in the 3D orientation.

The inset in Fig. \ref{fig:CH2O}(a) summarizes the foregoing discussion. Note that initially there is a small orientation along the $-Y$ direction, in accordance with the potential analysis.
However, due to its relatively small moments of inertia (see Table \ref{tab:Molecular-properties}), the  $\mathrm{CH_2O}$ molecule rotates relatively fast, resulting in a rapid change of orientation direction from $-Y$ to $Y$. Following the pulse, when the field-free 3D alignment appears at $t\simeq 0.2\,\mathrm{ps}$, the orientation of the dipole moment is along the $Y$ direction.

On the long time scale, in addition to the transient effect of 3D orientation, there is a residual orientation lasting long after the end of the pulse, see the inset in Fig. \ref{fig:CH2O}(b).
In contrast to the classically calculated orientation signal, the quantum one exhibits quantum beats \cite{Eberly1980,Parker1986,Averbukh1989,Felker1992Rotational,Robinett2004}. Despite the beats, the moving time average of this signal, like in the classical case, remains non-zero.
Related effects of long-lasting orientation were analyzed and implemented using other orientation schemes \cite{Milner2019Controlled,Tutunnikov2019Laser,Tutunnikov2020Observation,Xu2020,TutunnikovXu2020,xu2021longlasting}.
The existence of a non-zero time-averaged dipole signal relies on the combination of two factors: (i) the ability of symmetric/asymmetric-top molecules to precess about the conserved angular momentum vector and (ii) the symmetry breaking which is induced by the electromagnetic interaction. Here, the molecular $a$ axis ($\boldsymbol{\mu}\propto-\bm{a}$) performs a precession-like motion about the polarization direction of the SH field, contributing to the long-lasting dipole orientation.

The eigenstates of asymmetric-top molecules, $\ket{J\tau M}$ do not have a well-defined projection of the angular momentum on the molecular $a$ axis (unlike the eigenstates of symmetric-top molecules, $\ket{JKM}$). This leads to the decay of the long-lasting orientation and eventual change of its sign. However, as we show here, the time scale of this process exceeds the duration of the excitation pulse by orders of magnitude.

\begin{table*}[!t]
\caption{\label{tab:Molecular-properties-PPO}Calculated (using GAUSSIAN \cite{Frisch2016Gaussian}, method: CAM-B3LYP/aug-cc-pVTZ) molecular properties (in atomic units) of PPO: moments of inertia, elements of the dipole moment, polarizability tensor, and hyperpolarizability tensor in the frame of principal axes.
}
\begin{indented}
\item[]
\begin{tabular}{@{}ccccccc}\br
 Molecules                  & Moments of inertia            & Dipole
components           & \multicolumn{2}{l}{Polarizability components}                    & \multicolumn{2}{l}{Hyperpolarizability components}                    \\ \mr
\multirow{4}{*}{(\emph{R})-PPO} & \multirow{8}{*}{\begin{tabular}[c]{@{}c@{}}$I_{a}=180386$\\ \\ $I_{b}=493185$\\ \\ $I_{c}=553513$\end{tabular}} & \multirow{4}{2cm}{$\mu_{a}=0.380$\\$\mu_{b}=-0.682$\\$\mu_{c}= 0.192$}& \multirow{8}{1.8cm}{$\alpha_{aa}=45.63$ \\ $\alpha_{bb}=37.96$ \\ $\alpha_{cc}=37.87$ \\ $\alpha_{ab}=2.56$}  & \multirow{4}{1.6cm}{$\alpha_{ac}=0.85$\\ $\alpha_{bc}=0.65$} & \multirow{8}{2.1cm}{$\beta_{aaa}=32.18$\\$\beta_{aab}=-19.73$\\
$\beta_{abb}=3.12$\\$\beta_{acc}=8.61$\\$\beta_{bbb}=-17.77$\\
$\beta_{bcc}=-23.66$} & \multirow{3}{2.2cm}{$\beta_{aac}=10.79$\\$\beta_{abc}=-0.057$
\\$\beta_{bbc}=-2.38$\\$\beta_{ccc}=7.05$} \\
                    &                         &                     &                     &                     &                     \\
                    &                         &                     &                     &                     &                     \\
                    &                         &                     &                     &                     &                     \\ \cline{1-1} \cline{3-3} \cline{5-5} \cline{7-7}
\multirow{4}{*}{(\emph{S})-PPO} &                     & \multirow{4}{2cm}{$\mu_{a}=0.380$\\$\mu_{b}=-0.682$\\$\mu_{c}= -0.192$}    &                     & \multirow{4}{1.8cm}{$\alpha_{ac}=-0.85$ \\ $\alpha_{bc}=-0.65$} &                     & \multirow{4}{2.2cm}{$\beta_{aac}=-10.79$\\$\beta_{abc}=0.057$
\\$\beta_{bbc}=2.38$\\$\beta_{ccc}=-7.05$}  \\
                    &                         &                     &                     &                     &                     \\
                    &                         &                     &                     &                     &                     \\
                    &                         &                     &                     &                     &                     \\ \br
\end{tabular}
\end{indented}
\end{table*}

\subsection{Sulfur dioxide}  \label{sec:sulfur_dioxide}

\begin{figure}[!b]
\centering{}\includegraphics[width=\linewidth]{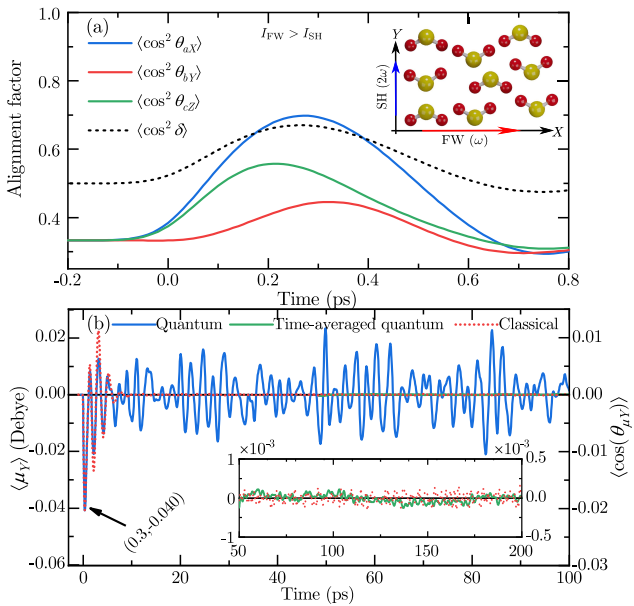}
\caption{(a) Classically calculated degrees of alignment and
(b) $Y$ projection of the dipole signal for $\mathrm{SO_2}$ molecule.
The conditions used are the same as in Fig. \ref{fig:CH2O}, except that $I_\mathrm{FW}=8\times 10^{13} \,\mathrm{W/cm^2}$ and $I_\mathrm{SH}=2\times 10^{13} \,\mathrm{W/cm^2}$. The overall degree is defined as $\braket{\cos^2\delta} = \left(1+\braket{\cos^2\theta_{aX}}+\braket{\cos^2\theta_{bY}}+\braket{\cos^2\theta_{cZ}}\right)/4$ (dashed black).
 \label{fig:SO2}}
\end{figure}

Figure \ref{fig:SO2}(a) shows that shortly after the end of the two-color pulse excitation, at $t\simeq 0.3\,\mathrm{ps}$, the $\mathrm{SO_2}$ molecules are 3D aligned as illustrated in the inset.
Here $I_\mathrm{FW}=8\times 10^{13} \,\mathrm{W/cm^2}$ and $I_\mathrm{SH}=2\times 10^{13} \,\mathrm{W/cm^2}$, so that the strong FW aligns the most polarizable $a$ axis [see oxygen atoms (red) in the inset of Fig. \ref{fig:SO2}(a)], while the weak SH field aligns the second most polarizable $b$ axis [see the sulfur atom (yellow)].
In addition, as shown in Fig. \ref{fig:SO2}(b), the dipole signal is along the $-Y$ direction at $t=0.3\,\mathrm{ps}$. This means that there are more molecular dipoles [see the sulfur atom (yellow)] pointing along the $-Y$ direction.

For the case considered here, when the FW is stronger than the SH, the field-polarizability potential [see Eq. \eqref{eq:potential_polarizability}]
has four global minima at $M_1=(0,\pi/2,\pi/2)$, $M_2= (0,\pi/2, 3\pi/2)$, $M_3=(\pi,\pi/2,\pi/2)$, $M_4= (\pi,\pi/2,3\pi/2)$. Like in the case of $\mathrm{CH_2O}$ molecule, the existence of these minima implies 3D alignment. Moreover, the total potential, including the field-hyperpolarizability interaction [see Eq. \eqref{eq:potential_SO2}] has only two global minima at $M_1$ and $M_4$. Since
$$U(M_1)\bm{\mu}^\mathrm{T}=U(M_4)\bm{\mu}^\mathrm{T}=(0,-0.7877,0)^\mathrm{T},$$
the two-color pulse kicks the dipole moment ($\boldsymbol{\mu}\propto-\bm{b}$) towards the $-Y$ direction. This results in 3D orientation as shown in the inset of Fig. \ref{fig:SO2}(a).

Furthermore, both the classical result and moving time average of the quantum signal show \emph{no} long-lasting dipole orientation, see the inset in Fig. \ref{fig:SO2}(b). As was mentioned previously (see also \cite{Xu2020}), one of the requirements for the long-lasting/persistent orientation is the ability of symmetric- and asymmetric-top molecules to precess about the conserved vector of angular momentum. Here, by precession or precession-like motion, we mean that the tip of one of the molecular axes moves on a closed trajectory around the conserved vector of angular momentum. Consequently, the precessing vectors have a constant sign projection along the direction of the angular momentum. In asymmetric-top molecules, such trajectories exist only for the two molecular axes with the lowest and highest moments of inertia, the $a$ and $c$ axes, respectively \cite{poinsot1851theorie,Goldstein2002Classical}. The projection of the intermediate axis, $b$ along the direction of angular momentum does not conserve its sign, resulting in a zero time average. In this sense, the $\mathrm{SO_2}$ molecule is a special case, as its dipole moment is along the intermediate axis of inertia, $\boldsymbol{\mu}\propto-\bm{b}$. The instability of the intermediate axis is a well-known fact \cite{Goldstein2002Classical}. It has non-trivial dynamical implications, called the tennis racket effect or the Dzhanibekov effect, which are still being studied \cite{Ashbaugh1991, VANDAMME201717,Marde2020}.

\begin{figure*}[!t]
\centering{}\includegraphics[width=\linewidth]{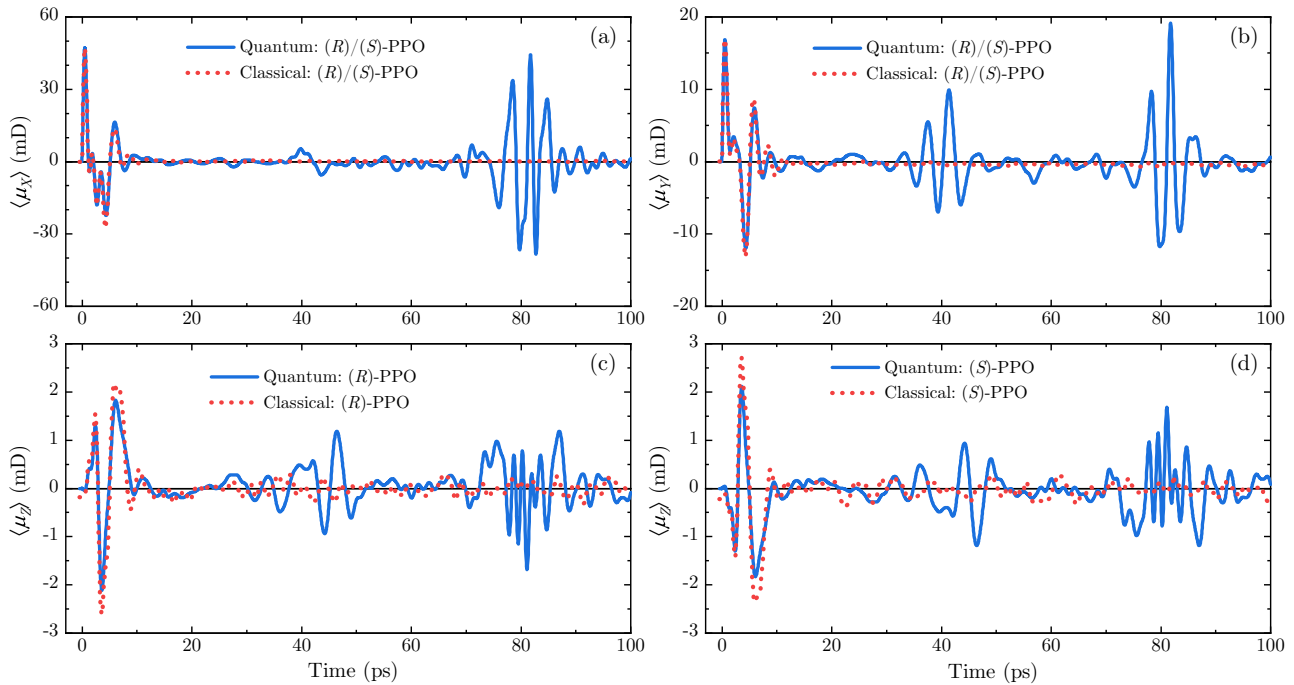}
\caption{Time-dependent projections of the dipole moment [measured in units of millidebye (mD)] on the laboratory axes for PPO molecules. The left- and right-handed enantiomers are denoted as (\emph{S})-PPO and (\emph{R})-PPO, respectively. Notice that the $Z$ projections of (\emph{R})-PPO (c) and (\emph{S})-PPO (d) are $\pi$ out of phase.
Solid blue and dotted red lines represent the results
of quantum and classical simulations, respectively.
Here the initial temperature is $T=5\,\mathrm{K}$, the field parameters used are the same as in Fig. \ref{fig:SO2}, except the angle between the polarizations of FW and SH, which here is $\phi_\mathrm{SH} = \pi/4$.
 \label{fig:PPO}}
\end{figure*}

\section{Enantioselective orientation effect}  \label{sec:enantioselective-orientation-effect}

In this section, we consider the orientation of chiral molecules excited by a two-color laser pulse. We use the propylene oxide molecule (PPO, $\mathrm{CH_{3}CHCH_{2}O}$), as a typical example. \Tref{tab:Molecular-properties-PPO} summarizes the molecular properties of the two enantiomers, (\emph{R})-PPO (right-handed) and  (\emph{S})-PPO (left-handed).
The two enantiomers are mirror images of each other and here we choose the $ab$-plane as the plane of reflection. The results are independent of the choice of the reflection plane \cite{Cotton1990Chemical}.

For (\emph{R})-PPO, the field-hyperpolarizability interaction potential induced by the orthogonally polarized two-color pulse (same as in Fig. \ref{fig:SO2}) has two global minima at $M_1^{(R)}\simeq(0.24\pi,0.25\pi,1.28\pi)$
and $M_2^{(R)}\simeq(0.76\pi,0.75\pi,0.28\pi)$. Accordingly,
\begin{align*}
U\left[M_1^{(R)}\right]\left[\bm{\mu}^{(R)}\right]^\mathrm{T}&\simeq(0.22, 0.76, -0.14)^\mathrm{T},\\
U\left[M_2^{(R)}\right]\left[\bm{\mu}^{(R)}\right]^\mathrm{T}&\simeq(-0.22, 0.76, 0.14)^\mathrm{T},
\end{align*}
where $\bm{\mu}^{(R)}$ is the dipole moment vector of the (\emph{R})-PPO molecule.
This shows that the two-color pulse kicks the dipole moment towards two directions, $(0.22, 0.76, -0.14)$ and $(-0.22, 0.76, 0.14)$ in the laboratory frame, such that the net dipole signal is along the $Y$ direction. The same is true for the case of (\emph{S})-PPO molecule.

In the case of chiral molecules (lacking any symmetry), 3D orientation requires not only 3D alignment and orientation of the dipole moment, but the orientation of all three molecular axes. 3D orientation of chiral molecules was studied using time-independent cross-polarized two-color laser fields in which the polarizations of the FW and SH fields are neither parallel nor orthogonal \cite{Takemoto2008}. Moreover, when the relative angle between the polarization is not an integer multiple of $\pi/2$, the induced orientation is enantioselective. As an example, we consider the case of $\phi_\mathrm{SH}=\pi/4$ [see Eq. \eqref{eq:electric-field-lab-frame}], namely the polarization of the FW is along the $X$ axis and the polarization of the SH is directed at an angle of $\pi/4$ with respect to the $X$ axis. For the molecular parameters of PPO, the angle $\phi_\mathrm{SH}=\pi/4$ was found to be close to optimal in terms of enantioselectivity.

Figure \ref{fig:PPO} shows the projections of dipole moment along the laboratory $X$, $Y$, and $Z$ axes as functions of time.
The two-color pulse induces orientation in the $X$ and $Y$ directions, and these orientations
are of the same sign for both enantiomers [see Figs. \ref{fig:PPO}(a) and \ref{fig:PPO}(b)].
Moreover, Figs. \ref{fig:PPO}(c) and \ref{fig:PPO}(d) show the appearance of the dipole signal along the laser propagation axis ($Z$ axis), $\braket{\mu_Z}(t)$. $\braket{\mu_Z}(t)$ is \emph{enantioselective}, namely it has an opposite sign for the two enantiomers.
A similar type of orientation was predicted in the case of chiral molecules excited by THz pulses with twisted polarization \cite{TutunnikovXu2020}.
In these cases, all three projections of the dipole moment, $\braket{\mu_X},\,\braket{\mu_Y},\,\braket{\mu_Z}$ are non-zero, and the projection along the field propagation direction is enantioselective.
In contrast, laser pulses with twisted polarization induce (enantioselective) orientation only along the laser propagation direction \cite{Yachmenev2016,Gershnabel2018,Tutunnikov2018,Milner2019Controlled,Tutunnikov2019Laser,Tutunnikov2020Observation}.
As was shown in the cases of laser/THz pulses with twisted polarization, the time scale of polarization twisting should be comparable with that of the molecular rotation. In contrast, here the demonstrated enantioselective orientation is caused by a single impulsive excitation.

The analysis of the interaction potential
shows that when $\phi_\mathrm{SH}=\pi/4$, the field-hyperpolarizability interaction potentials of (\emph{R})-PPO and (\emph{S})-PPO molecules have a \emph{single} global minimum at
\begin{align*}
M^{(R)}&\simeq(0, 0.32\pi,1.43\pi), \,
M^{(S)}&\simeq(0, 0.68\pi, 1.57\pi).
\end{align*}
The existence of a unique global minimum implies that, in principle, the chiral molecule can be fully oriented in space \cite{Takemoto2008}. In addition,
\begin{align*}
U\left[M_2^{(R)}\right]\left[\bm{\mu}^{(R)}\right]^\mathrm{T}&\simeq(0.50, 0.62, -0.09)^\mathrm{T}, \\
U\left[M_2^{(S)}\right]\left[\bm{\mu}^{(S)}\right]^\mathrm{T}&\simeq(0.50, 0.62, 0.09)^\mathrm{T},
\end{align*}
where $\bm{\mu}^{(S)}$ is the dipole moment vector of (\emph{S})-PPO molecule.
This means that the two-color field kicks the dipole moment towards a certain direction in space, such that the dipole projections on all three laboratory axes $X,Y,Z$ are non-zero.
The sign of the projection along the $Z$ direction is opposite for the two enantiomers, in agreement with the enantioselective orientation visible in Figs. \ref{fig:PPO}(c) and \ref{fig:PPO}(d).

In the case of non-chiral $\mathrm{CH_2O}$ and $\mathrm{SO_2}$ molecules, $\braket{\mu_Z}=0$ for all $\phi_\mathrm{SH}$ [see Eq. \eqref{eq:electric-field-lab-frame}]. As an example, we consider the $\mathrm{CH_2O}$ molecule and $\phi_\mathrm{SH}=\pi/4$. The field-hyperpolarizability interaction potential has two minima at
$M_1 \simeq(0.37\pi,\pi/2,\pi/2)$ and $M_2 \simeq(0.37\pi,\pi/2,3\pi/2)$, such that
$$U(M_{1})\bm{\mu}^\mathrm{T}=U(M_{2})\bm{\mu}^\mathrm{T}\simeq(-0.38, -0.87,0)^\mathrm{T},$$
which implies zero dipole orientation along the laser propagation direction.

\section{Conclusions}\label{sec:Conclusion}
We have investigated the orientation of asymmetric-top (including chiral) molecules excited by two-color femtosecond laser pulses.
In the case of planar asymmetric-top molecules, including formaldehyde and sulfur dioxide, the excitation by an orthogonally polarized two-color pulse leads to the 3D orientation. The orientation direction can be controlled by the relative angle between the polarizations of the fundamental wave and the second harmonic. The degree of orientation can be enhanced by optimizing the pulse parameters and by applying a sequence of multiple delayed pulses \cite{Averbukh2001,Averbukh2003,Averbukh2004,Bisgaard2004,Pinkham2007,Ren2014,Zhang2011}. The studied 3D molecular orientation may be useful e.g., in imaging of atomic motion using free-electron lasers and electron diffraction \cite{Kupper2014,Yang2016,Glownia2016}.
In addition, in the case of the formaldehyde molecule, the orientation lasts long after the end of the pulse.
This long-lasting orientation of the dipole moment may, potentially, be probed by even-harmonic generation and may be useful for enhancing the deflection of molecular beams in the presence of inhomogeneous electrostatic fields \cite{Kupper2012}.
Uniquely to chiral molecules, the induced orientation appears along the laser propagation direction when the polarizations of the fundamental wave and the second harmonic are neither parallel nor orthogonal. Moreover, the sign of the orientation is opposite for the two enantiomers. This enantioselective orientation may be useful for ultra-fast enantiomeric excess analysis and potentially also for the eventual separation of the two enantiomers.

\section*{Data availability statement}
The data that support the findings of this study are available upon reasonable request from the authors.


\ack{
I.A. gratefully acknowledges support by the Israel Science
Foundation (Grant No. 746/15). I.A. acknowledges support as the Patricia Elman
Bildner Professorial Chair. This research was made possible in
part by the historic generosity of the Harold Perlman Family.

}

\appendix

\begin{strip}
\section{Field-polarizability and field-hyperpolarizability interaction potentials}\label{app:potential}

The external field vector $\mathbf{E}$ can be expressed in the molecular reference frame with the help of the transformation matrix parametrized by the three Euler angles  $\phi,\,\theta,\,\chi$ [transpose of the matrix $U$ in Eq. \eqref{eq:transformation}].
Accordingly, the interaction potential [see Eq. \eref{eq:potential}] becomes a function
of $\phi,\,\theta,\,\chi$ as well.
For the orthogonally polarized two-color pulse, the field-polarizability
interaction of asymmetric-top molecule ($\mathrm{CH_2O}$ and $\mathrm{SO_2}$) is given by
\begin{footnotesize}
\begin{align}\label{eq:potential_polarizability}
V_{\alpha} =&-\frac{1}{4} \varepsilon_1^{2}\Bigg\{\Big[\cos(\theta) \cos (\phi) \cos (\chi)-\sin (\phi) \sin(\chi)\Big]^{2} \alpha_{bb}+\Big[\cos(\theta) \cos (\phi) \sin(\chi) + \sin(\phi)\cos(\chi) \Big]^{2} \alpha_{cc} + \sin^{2} (\theta)\cos^{2} (\phi)  \alpha_{aa}\Bigg\}\nonumber\\
&-\frac{1}{4} \varepsilon_2^{2}\Bigg\{\Big[\cos(\theta) \sin (\phi)\cos (\chi) +\cos (\phi) \sin (\chi)\Big]^{2} \alpha_{bb}+\Big[\cos (\theta) \sin(\phi) \sin(\chi) - \cos (\phi) \cos (\chi)\Big]^{2} \alpha_{cc}+\sin^{2}(\theta) \sin^{2} (\phi) \alpha_{aa}\Bigg\}.
\end{align}
\end{footnotesize}

The field-hyperpolarizability interaction potentials of $\mathrm{CH_2O}$ and $\mathrm{SO_2}$ are given by
\begin{small}
\begin{align}    \label{eq:potential_CH2O}
V_{\beta} =-\frac{\varepsilon_1^2\varepsilon_2}{64} \Bigg\{
&\Big(4\sin(\theta)\sin(\phi)\Big[6\cos^2(\theta)\cos^2(\phi)\cos^2(\chi)-[1+3\cos(2\phi)]\sin^2(\chi)\Big]+\Big[\cos(\phi)+3\cos(3\phi)\Big]\sin(2\theta)\sin(2\chi)\Big)\beta_{abb} \nonumber\\
+& \Big(4\sin(\theta)\sin(\phi)\Big[6\cos^2(\theta)\cos^2(\phi)\sin^2(\chi)-[1+3\cos(2\phi)]\cos^2(\chi)\Big]-\Big[\cos(\phi)+3\cos(3\phi)\Big]\sin(2\theta)\sin(2\chi)\Big)\beta_{acc}\nonumber\\
+&8\sin^3(\theta)\cos^2(\phi)\sin(\phi)\beta_{aaa}\Bigg\},
\end{align}
\end{small}
and \begin{small}
\begin{align}    \label{eq:potential_SO2}
V_{\beta} =-\frac{\varepsilon_1^2\varepsilon_2}{32}
\Bigg\{
&4\Big[\cos(\theta)\sin(\phi)\cos(\chi)+\cos(\phi)\sin(\chi)\Big]
\Big[\cos(\theta)\cos(\phi)\cos(\chi)-\sin(\phi)\sin(\chi)\Big]^2\beta_{bbb}\nonumber\\
+&\Big[\cos(\theta)\cos(\phi)\sin(\chi)+\sin(\phi)\cos(\chi)\Big]
\Big[3[1+\cos^2(\theta)]\sin(2\phi)\sin(2\chi)
-2\cos(\theta)[1+3\cos(2\phi)\cos(2\chi)]
\Big]\beta_{bcc}\nonumber\\
+&2\sin^2(\theta)\cos(\phi)\Big[3\cos(\theta)\sin(2\phi)\cos(\chi)+[3\cos(2\phi)-1]\sin(\chi)\Big]\beta_{aab}\Bigg\},
\end{align}
\end{small}
respectively.

\section{The transformation between laboratory and molecular frames}\label{app:transformation}

The transformation between the laboratory and molecular frames is given by $(X,Y,Z)^T=U(\phi,\theta,\chi)(b,c,a)^T$, where the transformation matrix reads
\begin{small}
\begin{align}\label{eq:transformation}
    U(\phi,\theta,\chi)
    =\begin{pmatrix}
    \cos(\phi) \cos(\theta) \cos(\chi) - \sin(\phi) \sin(\chi) & -\cos(\phi) \cos(\theta) \sin(\chi) - \sin(\phi) \cos(\chi) & \cos(\phi) \sin(\theta)\\
    \sin(\phi) \cos(\theta) \cos(\chi) + \cos(\phi) \sin(\chi) & -\sin(\phi) \cos(\theta) \sin(\chi) + \cos(\phi) \cos(\chi) & \sin(\phi) \sin(\theta)\\
    -\sin(\theta) \cos(\chi) & \sin(\theta) \sin(\chi) & \cos(\theta)
  \end{pmatrix}.
\end{align}
\end{small}

\end{strip}


\bibliography{references}

\end{document}